\documentclass[9pt,twocolumn,twoside]{osajnl}
\usepackage[justification=justified]{caption} 
\usepackage{siunitx}
\usepackage{amssymb}
\usepackage{amsfonts}
\usepackage{subcaption}
\usepackage{mathtools}
\usepackage{graphicx}

\journal{ol} 

\setboolean{shortarticle}{true}

\title{Artifacts in Optical Projection Tomography Due to Refractive-Index Mismatch: Model and Correction}

\author[1]{Yan Liu}
\author[1]{Jonathan Dong}
\author[2]{C\'edric Schmidt}
\author[1]{Aleix Boquet-Pujadas}
\author[2]{J\'er\^ome Extermann}
\author[1,*]{Michael Unser}

\affil[1]{Biomedical Imaging Group, \'Ecole polytechnique f\'ed\'erale Lausanne, Station 17, 1015 Lausanne, Switzerland}
\affil[2]{HEPIA/HES-SO, University of Applied Sciences of Western Switzerland, Rue de la Prairie 4, 1202 Geneva, Switzerland}

\affil[*]{Corresponding author: michael.unser@epfl.ch}

\dates{Compiled \today}

\doi{\url{https://doi.org/10.1364/OL.457144}}

\begin{abstract}
Optical projection tomography (OPT) is a powerful tool for 3D imaging of mesoscopic samples.
While it is able to achieve resolution of a few tens of microns over a sample volume of several cubic centimeters, the reconstructed images often suffer from artifacts caused by  inaccurate calibration.
In this work, we focus on the refractive-index mismatch between the sample and the surrounding medium.
We derive a 3D cone-beam forward model of OPT that approximates the effect of refractive-index mismatch. We then implement a fast and efficient reconstruction method to correct for the induced seagull-shaped artifacts on experimental images of fluorescent beads. 

\end{abstract}

\setboolean{displaycopyright}{true}

\begin{document}

\maketitle



\def\pixel{px}
Since its invention by J. Sharpe in 2002  \cite{Sharpe:02, Sharpe:04}, optical projection tomography (OPT) has become a powerful tool to obtain three-dimensional images of biological tissues at the mesoscopic scale \cite{Darrell:08, Schmidt:21, Nguyen:17}. 
Due to its significant penetration depth \cite{JochenBirk:11}, it can image whole animals at a resolution of a few tens of microns \cite{Schmidt:21} with molecular specificity \cite{Darrell:08, Schmidt:21}. 
Often referred to as the optical analog of X-ray computed tomography \cite{Kak:01}, OPT works with non-ionizing light, thereby minimizing radiation damage and facilitating studies of a broad range of diseases that include Alzheimer's \cite{Lindsey:17} and gastrointestinal pathologies \cite{Schmidt:21}. 

Regrettably, one frequently observes various artifacts in OPT images for a range of different reasons. 
For instance, mechanical errors in the imaging system may result in a misaligned rotation center that will generate double-edged or circular artifacts in the reconstruction \cite{Walls:05, michalek:15, Torres:21}. 
Unstabilities such as fluctuations of illumination and variations of the sensitivity of the detector will cause smear and ring artifacts \cite{Walls:05}. 
Finally, deviations from the physical model may arise such as the finite depth-of-field of the imaging system \cite{Sharpe:04} or a refraction that perturbs the propagation of light \cite{JochenBirk:11, Antonopoulos:14}. 
All these artifacts come from a mismatch between the OPT model and the actual physical realization. 
They strongly degrade the final resolution and quality of the reconstructed OPT images but can be computationally corrected once they have been properly identified \cite{Walls:05, michalek:15}. 

During an OPT experiment, the sample is typically embedded in a cylindrical gel phantom that is immersed in an index-matching liquid \cite{Doran:01, Islam:03, Sharpe:04}, in an attempt to limit the refraction of light at the interface of the agarose gel and the embedding liquid \cite{Doran:01, Islam:03, Haidekker:06, marcos:20}. 
Despite their best effort, OPT practitioners still observe seagull-shaped artifacts suspected to be caused by a residual mismatch in the refractive indices (RI) between the sample and its embedding medium \cite{Schmidt:21}. 
A previous study had already reported similar artifacts, mostly located around the edge of the sample, and proposed a computational method to compensate for the RI mismatch \cite{JochenBirk:11}. However, as this method requires volumetric alignment, it remains computationally expensive. 

In this work, we propose instead a simple ray-optics model to remove artifacts induced by the RI mismatch between the cylindrical gel and the surrounding liquid.
Our contributions are twofold: study of this type of RI mismatch and its interpretation as a virtual cylindrical lens;
approximation of the virtual lensing effect with the help of a cone-beam model. 
We discuss the validity of our model and point out the importance of the displacement of the rotating cylinder. 
We then design a computationally-efficient algorithm that leverages our model to correct the artifacts on experimental images of fluorescent beads.

\begin{figure*}[htb]
\begin{minipage}[b]{0.47\linewidth}
  \centerline{\includegraphics[width=\linewidth]{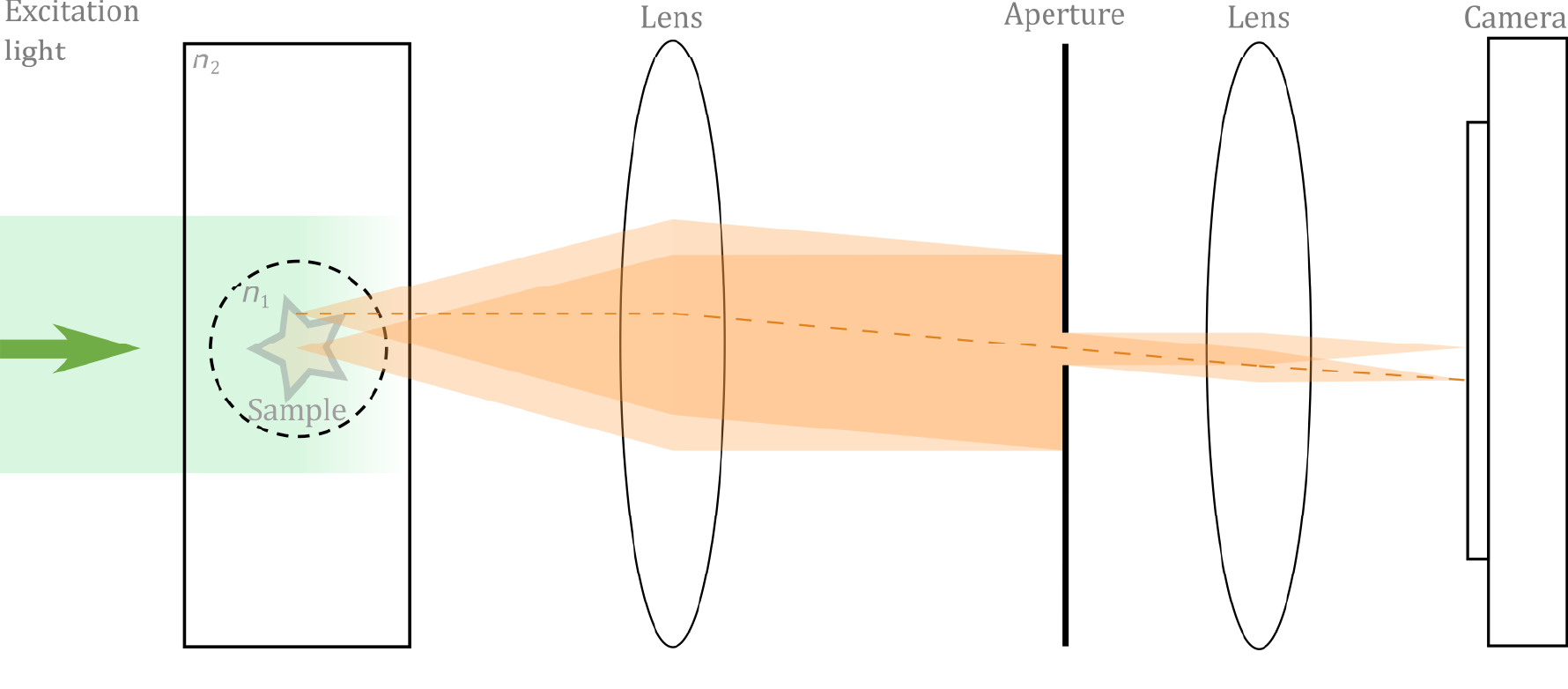}}
\end{minipage}
\hfill
\begin{minipage}[b]{0.47\linewidth}
    \includegraphics[width=\linewidth]{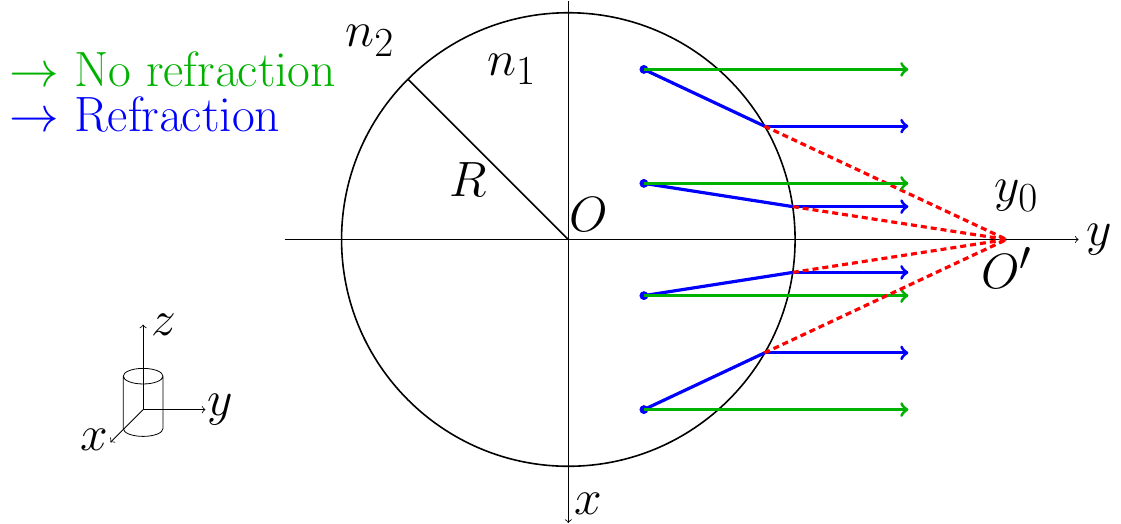}
\end{minipage}
\caption{OPT geometry in the ideal case and in the non-ideal case when the refractive indices do not match. (Left) Emission OPT  viewed from the top. The sample, fixed in agarose gel of RI $n_1$, is placed inside a glass container filled with an embedding liquid of RI $n_2$. The fluorophores inside the sample emit light that propagates through the object-space telecentric lens system and gets recorded by the camera. More details about the experimental system can be found in \cite{Schmidt:21}. (Right) Ideal case (green) and RI mismatch (blue). Only rays parallel to the optical axis are considered due to the low-NA fluorescence collection in OPT. The RI $n_1$  of the agarose differs from the RI $n_2$ of the embedding medium. This creates a lensing effect, with a virtual focus at position $O'$ on the optical axis.
}
\label{fig:fig1}
\end{figure*}
In an OPT experiment, the sample is rotated to a series of angular positions to produce a set of 2D projections.
Before the experiment, the rotation axis is carefully adjusted to be perpendicular to the optical axis, so that the projection data of one slice of the volume are recorded in a row of pixels on the detector \cite{Sharpe:04}. 
When there is no refraction, diffraction, scattering, or defocusing, the emitted light propagates through the sample, the index-matching liquid, and the telecentric lens system in parallel straight lines (see Fig. \ref{fig:fig1}). 
Accordingly, the standard forward model for OPT is the Radon transform \cite{Natterer:01, Koljonen:19}, which in 3D is typically inverted in a slice-by-slice fashion using the filtered backprojection (FBP) algorithm \cite{Sharpe:04, Natterer:01}.

In reality, when the RI of the sample does not match the RI of the embedding liquid, the path along which the light propagates is altered so that the parallel-beam assumption no longer holds. 
Consider a 2D transverse slice of the sample in the liquid bath as depicted in Fig. \ref{fig:fig1}. 
We denote the refractive index of the agarose gel and the liquid bath as $n_1$ and $n_2$, respectively, and the radius of the cylinder as $R$.
When $n_1=n_2$, the light follows parallel straight lines as it travels through the sample, the liquid bath, and finally the flat wall of the cuvette (see the green lines in Fig. \ref{fig:fig1}).
When $n_1 < n_2$, the light will be refracted at the interface that separates the agarose cylinder from the liquid bath. 
The propagation of the refracted light is illustrated by the blue lines in Fig. \ref{fig:fig1}. 
The extended incident rays (blue lines inside the cylinder in Fig. \ref{fig:fig1}) intersect at a point $O'$ on the optical axis (dotted red lines in Fig. \ref{fig:fig1}). 
Hence, the front wall (close to the detector side) of the cylinder can be thought of as a diverging lens whose virtual focus $O'$ is on the optical axis. 
The rays inside the cylinder can then be approximated by a fan beam the origin of which coincides with the virtual focus.

Such a picture is based on two assumptions, common in ray optics, to keep our model computationally tractable. 
First, we make the paraxial assumption; namely, that the point emitters are close to the optical axis. 
Second, we consider the difference in refractive indices to be small ($\Delta n = |n_2-n_1| \ll n_1$), so that the resulting origin of the fan beam is far away from the cylinder.
Under these assumptions, the location $y_0$ of the origin $O'$ of the converging beam depends on the radius $R$ of the cylinder and the refractive indices ($n_1,n_2$) according to 
\begin{equation}
    y_0=\frac{n_1}{n_2-n_1}R.
    \label{eq:origin}
\end{equation}
A detailed derivation of \eqref{eq:origin}  based on the ray geometry of Fig. \ref{fig:fig1} is provided in the supplement.
Note that the unitless quantity $y_0/R$ is not affected by the magnification of the imaging system.

Therefore, when $n_1 \neq n_2$, this physical model differs from the one typically used for reconstruction with a parallel-beam geometry.
Yet, the proposed virtual-lens model is a better approximation of the actual physics. We are going to show that it is capable of reproducing and eventually correcting for some of the reconstruction artifacts.
This 2D analysis corresponds to a fan-beam model; we can extend it to a cone-beam model when the height of the volume is sufficiently small.
Cone-beam geometries are common in other tomographic applications, which allows us to leverage efficient implementations and reconstruction algorithms.

\begin{figure}[!b]
\begin{minipage}[b]{1.0\linewidth}
  \centering
  \centerline{\includegraphics[width=9cm]{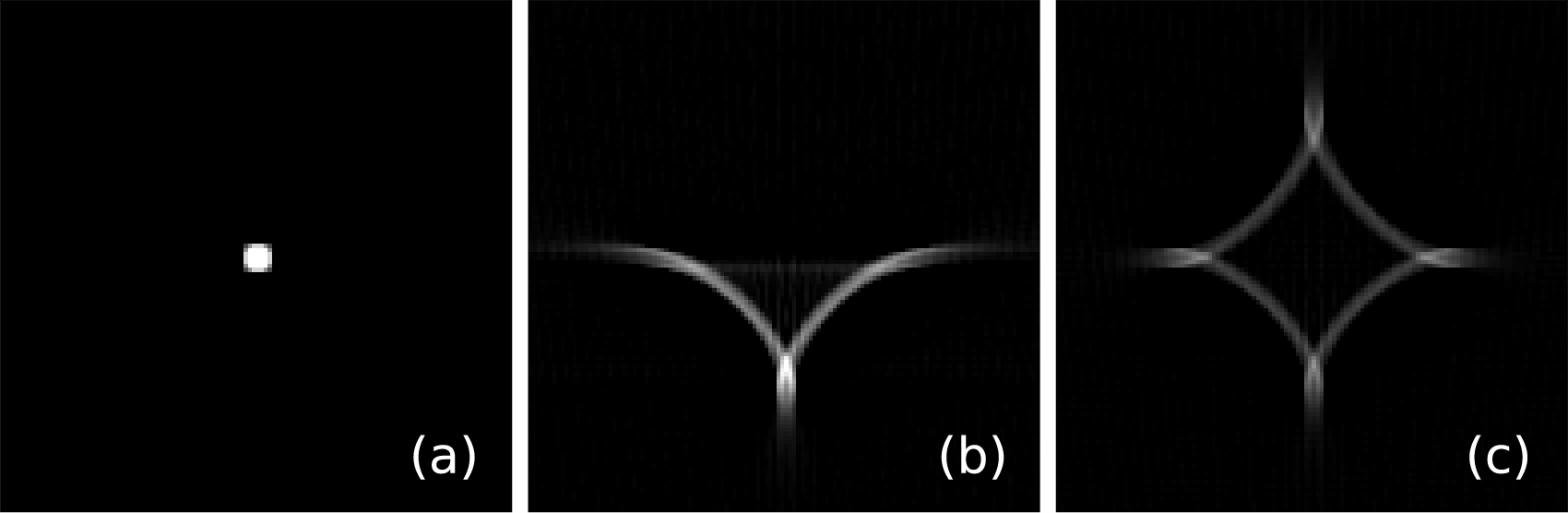}}
\end{minipage}
\caption{``Seagull'' artifacts due to mismatched refractive indices with rotation angles $[0,\pi)$ (b) and $[0,2\pi)$ (c), compared to the ground-truth fluorescent bead (a).}
\label{fig:simu}
\end{figure}

While \eqref{eq:origin} provides a good approximation of the beam origin in an idealised setting, two additional challenges arise in practice. First, the RI of the agarose gel evolves to match that of the surrounding liquid (e.g., BABB) due to molecular diffusion. 
This process may extend over the standard clearing protocol time, making it difficult to measure the value of $n_1$ and thus to calculate the origin using \eqref{eq:origin}. Second, there is another physical effect to consider: the cylinder is displaced as it rotates because of an experimental offset between the axis of rotation and the axis of the cylinder. 
Despite their best effort, instrumentalists have not yet been able to calibrate their setups accurately enough to avoid this offset \cite{Koljonen:19, Walls:07, Koskela:19}. For example, the seagull artifacts in \cite{Schmidt:21} (cf. Fig. \ref{fig:simu}f therein) can be reduced by decreasing the offset between the axis of the cylinder and the axis of rotation. In the supplement, we show that the cone-beam geometry remains a good approximation under these conditions, but with an origin that is closer to the sample than \eqref{eq:origin}.

To tackle both challenges at once, we propose a method to automatically determine the optimal origin of the cone beam by minimizing an heuristic loss metric.
More details are provided in the section presenting experiments on real data.  
\begin{figure*}[!t]
\begin{minipage}[b]{0.66\linewidth}
  \centerline{\includegraphics[width=\textwidth]{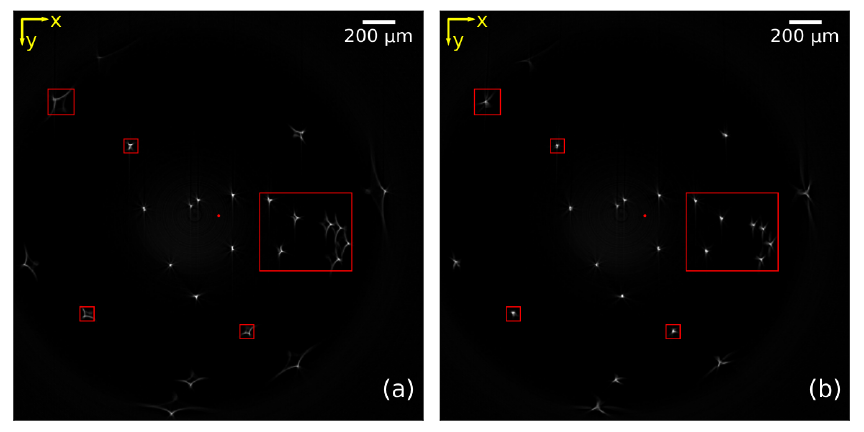}}
\end{minipage}
\begin{minipage}[b]{0.33\linewidth}
  \centerline{\includegraphics[width=\textwidth]{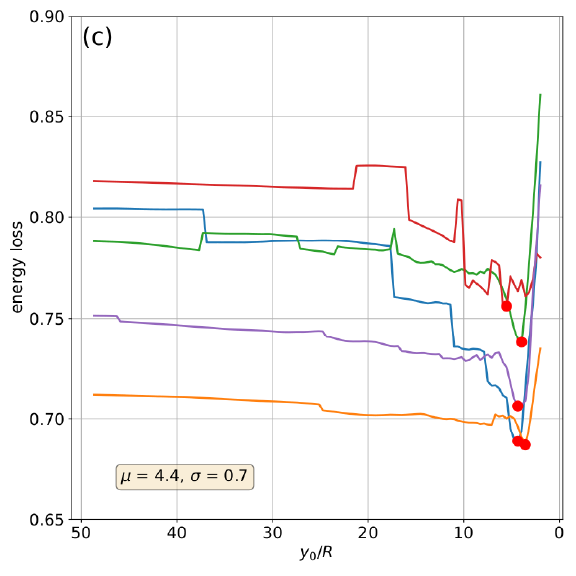}}
\end{minipage}

\centering
\begin{minipage}[b]{0.53\linewidth}
\centering
  \centerline{\includegraphics[width=\textwidth]{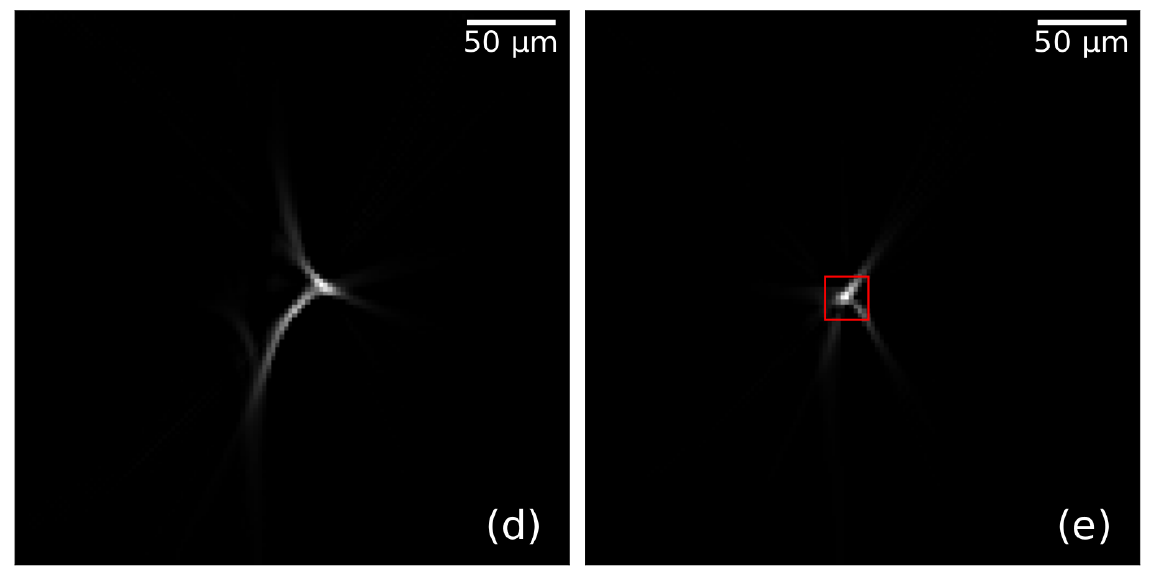}}
\end{minipage}
\centering
\begin{minipage}[b]{0.43\linewidth}
  \centerline{\includegraphics[width=.8\textwidth, height=4.9cm]{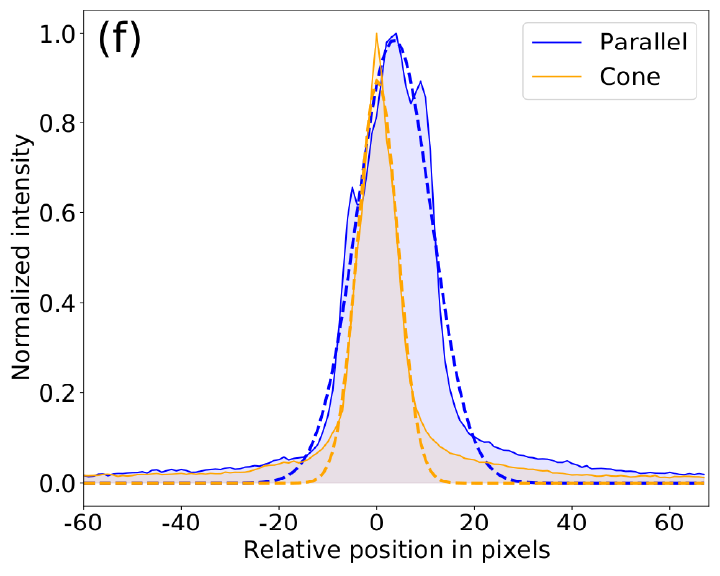}}
\end{minipage}

\caption{Reconstructions of a 3D volume of fluorescent beads integrated along the $z-$axis using a parallel-beam model (a) and a cone-beam model (b). (c): Energy loss of 5 beads over a range of origin guesses. The red point on each curve indicates the minimum. Same slice of the 3D reconstructions of a bead using parallel-beam (d) and cone-beam (e) models. The square in (e) is the region $\Omega_{\mathrm{in}}$ used to compute the energy loss. (f): Intensity profiles along the direction of maximum variance of the bead in (d) and (e). The blue and orange lines represent the intensity profiles for the parallel-beam and cone-beam models, respectively. The dashed blue and orange lines represent the fitted Gaussian point-spread functions.}
\label{fig:result}
\end{figure*}
To validate our model of the RI mismatch, we simulate the setup of a real emission OPT experiment for the imaging of fluorescent beads.  
We use the cone-beam model to generate projections with an origin position $y_0$ defined by $y_0/R=5$. We then reconstruct with the FBP algorithm. 
Our phantom consists of 13 beads, each of unit intensity and of size $(5\times5)$\pixel$\,$ (one pixel is 2.45 $\mu$m) embedded in a slice of a cylinder with radius $R=1024$\pixel$\,$ contained in a square cuvette of length $2048$\pixel. 
Moreover, it is important to consider that the depth of field in OPT usually covers only half of the sample, in contrast with X-ray tomography.
To approximate the out-of-focus phenomenon of the other half, we implement rotation angles over both $[0, \pi)$ and $[0,2\pi)$.

All simulations have been performed using the Tomosipo toolbox \cite{Hendriksen:21}, which is a wrapper for the tomographic-reconstruction library ASTRA \cite{vanAarle:16}.
We have adapted a cone-beam geometry as this modality is common in tomography \cite{Grass:99} and has already been implemented efficiently in Tomosipo.
It enables us to accelerate the computations on CPU and GPU and to use the already implemented Feldkamp-Davis-Kress (FDK) algorithm for cone-beam reconstructions \cite{Feldkamp:84} in stacks of slices of size 2 to obtain the 3D volume.
This is crucial in OPT because a time- and memory-efficient reconstruction algorithm is required to perform the high-resolution imaging of mesoscopic samples and to process huge datasets ($1200$ projections of $(2048 \times 2048)$~\pixel $\,$ each for the experiment presented below).
Accurate models such as \cite{JochenBirk:11} hit computational bottlenecks and become infeasible for temporal and spatial imaging at high-resolution.

We show in Fig. \ref{fig:simu} (a) a vertical section of the  reconstruction of a bead sitting at the exact center of the image.
This ground-truth image is free of artifacts.
As the distance of the bead to the center of the image increases, the seagull artifacts become more pronounced, which is reflected in the size of their ``wings''.
We give a closeup view of a non-centered bead with strong seagull artifacts in Fig. \ref{fig:simu} (b-c). More precisely, its reconstructions using the rotation angles in $[0,\pi)$ and $[0,2\pi)$ are shown in Fig. \ref{fig:simu} (b) and (c), respectively.
The magnified ground truth image is provided in Fig. \ref{fig:simu} (a) as a reference. In agreement with our model, the FBP-based reconstruction of a mismatched experiment introduces similar artifacts to those observed in real data.

Next, we show how to mitigate the ``seagull'' artifacts on experimental data of fluorescent beads.
Fluorescent FITC-labeled 1 $\mu$m microspheres based on melamine resin (90305-5ML-F, Sigma) were used as specimen, in accordance with the preparation procedure described in \cite{Schmidt:21}. Approximately 0.01~mL of particle solution was mixed to 10~mL of 1.5\% agarose before molding to achieve a nearly colloidal solution. The acquisition was performed shortly after preparation to avoid fluorescence quenching of the signal. 
The RI $n_2$ of BABB in our experiment is 1.56. 
The RI $n_1$ of the agarose changes over time, but the mismatch between the two materials is initially of the order of 0.1. 

Before reconstruction, we first compensate for the misaligned rotation center.
However, because the fluorescent beads are very sparse compared to an actual biological sample, they carry too little information for one to apply the standard method based on the maximum variance of a reconstructed slice \cite{Walls:07, Torres:21, Schmidt:21}. 
Instead, we perform a grid search and determine the best estimate of the center of rotation when the ``circle''-shaped artifacts caused by the wrong rotation center vanish. 

As discussed before, another step is to determine the optimal origin of the cone beam. We do this automatically by minimizing the loss metric
\begin{equation}
    E_{\mathrm{loss}}= 1 - \frac{\sum_{k\in \Omega_{\mathrm{in}}} I_k}{\sum_{k\in \Omega_{\mathrm{total}}}I_k},
    \label{eq:energy-loss}
\end{equation}
where $\Omega_{\mathrm{total}}$ is a square region that fully encloses the whole ``seagull'' artifact, $\Omega_{\mathrm{in}}$ is a smaller square region enclosing only the central part of a bead (see Fig. \ref{fig:result} (b), (d), and (f)), and $I_k$ denotes the intensity at pixel $k$.
The quantity $E_{\mathrm{loss}}$ hence indicates how concentrated the distribution of intensity is around the center of a bead. 
The smaller $E_{\mathrm{loss}}$ is, the less energy is spread outside the true location of the bead, and the better the reconstruction is. 
We provide in Fig. \ref{fig:result} the energy loss of five different beads at varied locations in the field of view.
We observe that the energy loss behaves similarly for all five beads and achieves its minimum at about the same value of each curve. 
This means that the optimal origin guesses are consistent over several beads, which confirms the robustness of the proposed automatic calibration. 
We thus adopt the average value $\mu=y_0/R= 4.4$ as the optimal origin $y_0$ for all the beads in the reconstruction. Compared to $\mu$, the theoretical value of $y_0/R=15$ derived from \eqref{eq:origin} highlights the importance of performing this automatic calibration to account for the origin shift discussed in the supplement.
The 3D reconstruction of a $2048\times2048\times1024$ volume and grid search to find the optimal origin over 200 points for one bead take 95 seconds and 108 seconds on CPU, respectively.

In Fig. \ref{fig:result}, we compare the reconstructions using the parallel-beam model or our approach. We can observe the clear radial dependence of the size of the seagull artifacts. Our proposed cone-beam model successfully mitigates the artifacts, especially for the beads identified by boxes. 
To quantify the improvement of our method, we provide the intensity profiles along the direction of maximal variance of the uncorrected and corrected reconstructions in Fig. \ref{fig:result} (f). 
The full-width half-maximum along the direction of maximum variance, calculated based on the Gaussian fit, is 47~$\mu$m and 22~$\mu$m for Fig.~  \ref{fig:result} (d) and (e), respectively.
This means that our corrections lead to a 52\% increase in resolution.


To conclude, this work models mismatching refractive indices in emission OPT as a virtual lens. 
Using the corresponding cone-beam model, we can reproduce in simulations the ``seagull'' artifacts observed in real emission OPT experiments. 
To achieve best performance in practice, we propose a fast approach that automatically determines the optimal origin of the cone beam. We perform reconstructions based on the cone-beam model and validate our methods on experimental data of fluorescent beads. 
Our results show that the cone-beam model, with an optimized origin, successfully mitigates the seagull-shaped artifacts and increases the resolution of the reconstruction in the directions perpendicular to the optical axis.
Future research directions include the design of a better model of the finite depth-of-field and of the rotating cylinder. 
This work could be extended to transmission OPT by taking into account the RI mismatch on both sides of the sample.
Finally, cone-beam reconstructions could be applied to non-pointwise samples.

\section*{Funding}

Y.L., J.D., and M.U. acknowledge funding from European Research Council (ERC) under the European Union’s Horizon 2020 research and innovation programme (Grant Agreement No. 692726 GlobalBioIm). C.S. and J.E. acknowledge funding from Innosuisse - Schweizerische Agentur für Innovationsförderung (31434.1 IP-ICT) and HESSO (P2 - in-vitro TBI).

\section*{Acknowledgments}

We would like to thank Thanh-an Pham for useful discussions and Eric Sinner for his help in making the figures.

\section*{Disclosures}
\noindent The authors declare no conflicts of interest.

\section*{Data availability}
Data underlying the results presented in this paper are available in \cite{datalink}.

\bibliography{biblio.bib}

\begin{thebibliography}{10}
\newcommand{\enquote}[1]{``#1''}

\bibitem{Sharpe:02}
J.~Sharpe, U.~Ahlgren, P.~Perry, B.~Hill, A.~Ross, J.~Hecksher-Sørensen,
  R.~Baldock, and D.~Davidson, \enquote{Optical projection tomography as a tool
  for 3d microscopy and gene expression studies,}
  {\protect\JournalTitle{Science (New York, N.Y.)}} \textbf{296}, 541--5
  (2002).

\bibitem{Sharpe:04}
J.~Sharpe, \enquote{Optical projection tomography,}
  {\protect\JournalTitle{Annual Review of Biomedical Engineering}} \textbf{6},
  209--28 (2004).

\bibitem{Darrell:08}
A.~Darrell, H.~Meyer, U.~Birk, K.~Marias, M.~Brady, and J.~Ripoll,
  \enquote{Maximum likelihood reconstruction for fluorescence optical
  projection tomography,} in \emph{2008 8th IEEE International Conference on
  BioInformatics and BioEngineering,}  (2008), pp. 1--6.

\bibitem{Schmidt:21}
C.~Schmidt, A.~L. Planchette, D.~Nguyen, G.~Giardina, Y.~Neuenschwander, M.~D.
  Franco, A.~Mylonas, A.~C. Descloux, E.~Pomarico, A.~Radenovic, and
  J.~Extermann, \enquote{High resolution optical projection tomography platform
  for multispectral imaging of the mouse gut,} {\protect\JournalTitle{Biomed.
  Opt. Express}} \textbf{12}, 3619--3629 (2021).

\bibitem{Nguyen:17}
D.~Nguyen, P.~J. Marchand, A.~L. Planchette, J.~Nilsson, M.~Sison,
  J.~Extermann, A.~Lopez, M.~Sylwestrzak, J.~Sordet-Dessimoz,
  A.~Schmidt-Christensen, D.~Holmberg, D.~V.~D. Ville, and T.~Lasser,
  \enquote{Optical projection tomography for rapid whole mouse brain imaging,}
  {\protect\JournalTitle{Biomed. Opt. Express}} \textbf{8}, 5637--5650 (2017).

\bibitem{JochenBirk:11}
U.~J. Birk, A.~Darrell, N.~Konstantinides, A.~Sarasa-Renedo, and J.~Ripoll,
  \enquote{Improved reconstructions and generalized filtered back projection
  for optical projection tomography,} {\protect\JournalTitle{Appl. Opt.}}
  \textbf{50}, 392--398 (2011).

\bibitem{Kak:01}
A.~C. Kak and M.~Slaney, \emph{Principles of Computerized Tomographic Imaging}
  (Society for Industrial and Applied Mathematics, 2001).

\bibitem{Lindsey:17}
B.~W. Lindsey and J.~Kaslin, \enquote{Optical projection tomography as a novel
  method to visualize and quantitate whole-brain patterns of cell proliferation
  in the adult zebrafish brain,} {\protect\JournalTitle{Zebrafish}}
  \textbf{14}, 574--577 (2017). PMID: 28296621.

\bibitem{Walls:05}
J.~R. Walls, J.~G. Sled, J.~Sharpe, and R.~M. Henkelman, \enquote{Correction of
  artefacts in optical projection tomography,} {\protect\JournalTitle{Physics
  in Medicine and Biology}} \textbf{50}, 4645--4665 (2005).

\bibitem{michalek:15}
J.~Michalek, \enquote{Total variation-based reduction of streak artifacts, ring
  artifacts and noise in 3d reconstruction from optical projection tomography,}
  {\protect\JournalTitle{Microscopy and Microanalysis}} \textbf{21},
  1602–1615 (2015).

\bibitem{Torres:21}
V.~C. Torres, C.~Li, W.~Zhou, J.~G. Brankov, and K.~M. Tichauer,
  \enquote{Characterization of an angular domain fluorescence optical
  projection tomography system for mesoscopic lymph node imaging,}
  {\protect\JournalTitle{Appl. Opt.}} \textbf{60}, 135--146 (2021).

\bibitem{Antonopoulos:14}
G.~C. Antonopoulos, D.~Pscheniza, R.-A. Lorbeer, M.~Heidrich, K.~Schwanke,
  R.~Zweigerdt, T.~Ripken, and H.~Meyer, \enquote{{Correction of image
  artifacts caused by refractive index gradients in scanning laser optical
  tomography},} {\protect\JournalTitle{Three-Dimensional and Multidimensional
  Microscopy: Image Acquisition and Processing XXI}} \textbf{8949}, 21 -- 26
  (2014).

\bibitem{Doran:01}
S.~Doran, K.~Koerkamp, M.~Bero, P.~Jenneson, E.~Morton, and W.~Gilboy,
  \enquote{A ccd-based optical ct scanner for high-resolution 3d imaging of
  radiation dose distributions: Equipment specifications, optical simulations
  and preliminary results,} {\protect\JournalTitle{Physics in Medicine and
  Biology}} \textbf{46}, 3191--213 (2002).

\bibitem{Islam:03}
K.~Islam, J.~Dempsey, M.~Ranade, M.~Maryanski, and D.~Low, \enquote{Initial
  evaluation of commercial optical ct-based 3d gel dosimeter,}
  {\protect\JournalTitle{Medical Physics}} \textbf{30}, 2159--68 (2003).

\bibitem{Haidekker:06}
M.~Haidekker, \enquote{Optical transillumination tomography with tolerance
  against refraction mismatch,} {\protect\JournalTitle{Computer Methods and
  Programs in Biomedicine}} \textbf{80}, 225--35 (2006).

\bibitem{marcos:20}
A.~Marcos-Vidal and J.~Ripoll, \enquote{Recent advances in optical tomography
  in low scattering media,} {\protect\JournalTitle{Optics and Lasers in
  Engineering}} \textbf{135}, 106191 (2020).

\bibitem{Natterer:01}
F.~Natterer, \emph{The Mathematics of Computerized Tomography} (Society for
  Industrial and Applied Mathematics, USA, 2001).

\bibitem{Koljonen:19}
V.~Koljonen, O.~Koskela, T.~Montonen, A.~Rezaei, B.~Belay, E.~Figueiras,
  J.~Hyttinen, and S.~Pursiainen, \enquote{A mathematical model and iterative
  inversion for fluorescent optical projection tomography,}
  {\protect\JournalTitle{Physics in Medicine and Biology}} \textbf{64} (2019).

\bibitem{Walls:07}
J.~R. Walls, J.~G. Sled, J.~Sharpe, and R.~M. Henkelman, \enquote{Resolution
  improvement in emission optical projection tomography,}
  {\protect\JournalTitle{Physics in Medicine and Biology}} \textbf{52},
  2775--2790 (2007).

\bibitem{Koskela:19}
O.~Koskela, T.~Montonen, B.~Belay, E.~Figueiras, S.~Pursiainen, and
  J.~Hyttinen, \enquote{Gaussian light model in brightfield optical projection
  tomography,} {\protect\JournalTitle{Scientific Reports}} \textbf{9} (2019).

\bibitem{Hendriksen:21}
A.~Hendriksen, D.~Schut, W.~J. Palenstijn, N.~Viganò, J.~Kim, D.~Pelt, T.~van
  Leeuwen, and K.~J. Batenburg, \enquote{Tomosipo: Fast, flexible, and
  convenient {3D} tomography for complex scanning geometries in {Python},}
  {\protect\JournalTitle{Optics Express}}  (2021).

\bibitem{vanAarle:16}
W.~van Aarle, W.~J. Palenstijn, J.~Cant, E.~Janssens, F.~Bleichrodt,
  A.~Dabravolski, J.~D. Beenhouwer, K.~J. Batenburg, and J.~Sijbers,
  \enquote{Fast and flexible x-ray tomography using the astra toolbox,}
  {\protect\JournalTitle{Opt. Express}} \textbf{24}, 25129--25147 (2016).

\bibitem{Grass:99}
M.~Grass, T.~Köhler, and R.~Proksa, \enquote{3d cone-beam {CT} reconstruction
  for circular trajectories,} {\protect\JournalTitle{Physics in Medicine and
  Biology}} \textbf{45}, 329--347 (1999).

\bibitem{Feldkamp:84}
L.~A. Feldkamp, L.~C. Davis, and J.~W. Kress, \enquote{Practical cone-beam
  algorithm,} {\protect\JournalTitle{J. Opt. Soc. Am. A}} \textbf{1}, 612--619
  (1984).

\bibitem{datalink}
\url{https://doi.org/10.5281/zenodo.6141020}.

\end{thebibliography}
\bibliographyfullrefs{biblio.bib}
\clearpage
\appendix
\section{Derivation of the Cone-Beam Equivalence}
\begin{figure}[tb]
    \centering
    \includegraphics[width=\linewidth]{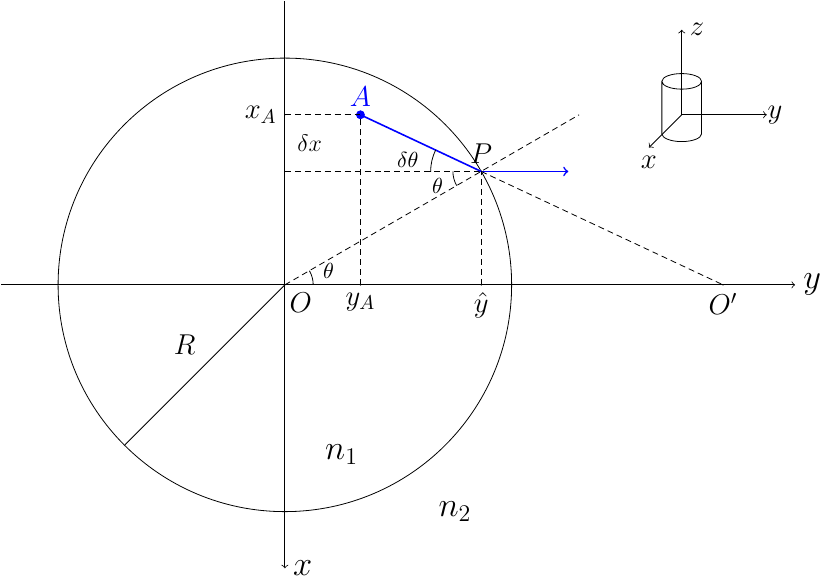}
    \caption{Geometry of light refraction at the interface of the cylinder and the embedding medium. The refractive indices inside and outside the circle are $n_1$ and $n_2$, respectively.}
\label{fig:geometry}
\end{figure}

To calculate the location of the virtual focus of the diverging lens (i.e., the origin of the cone beam), we focus on the 2D transverse slice of the cylinder shown in Fig. \ref{fig:geometry}).

The main quantities that will be useful are the following:
\begin{itemize}
    \item $R$ the radius of the disk
    \item $A (x_A, y_a)$ the position of emitter
    \item $P (\hat x, \hat y)$ the incident point on the surface of the cylinder
    \item $\delta x$ the vertical displacement between the emitter and $P$
    \item $O'$ the intersection of the incident ray with the $y$-axis
    \item $\theta$ the refracted angle such that the refracted ray is parallel to the optical axis
    \item $\theta + \delta \theta$ the incident angle
    \item $\Delta n=(n_2-n_1)$, assumed to be small with respect to $n_1$
\end{itemize}

Snell's law states that
\begin{equation}
    n_2\sin\theta=n_1\sin(\theta+\delta\theta)\approx n_1(\sin\theta+\delta\theta\cos\theta),
    \label{eq:snell}
\end{equation}
where we performed a first-order Taylor expansion.
\eqref{eq:snell} implies that 
\begin{equation}
    \delta\theta=\frac{n_2-n_1}{n_1}\tan\theta.
    \label{eq:deltatheta}
\end{equation}
Because $\delta\theta\ll1$, in the right triangle with longest side AP, one has that
\begin{equation}
    \delta\theta\approx\tan\delta\theta=\frac{\delta x}{\hat y-y_A},
    \label{eq:deltax}
\end{equation}
with $\hat y = \sqrt{R^2-(x_A-\delta x)^2} = \sqrt{R^2 - x_A^2} + \mathcal{O}(\Delta n)$.
Combining \eqref{eq:deltax} and \eqref{eq:deltatheta}, we obtain that
\begin{equation}
    \delta x =\left(\hat y-y_A\right)\frac{\Delta n}{n_1}\tan\theta.
    \label{eq:formula1}
\end{equation}
We also have that
\begin{equation}
    \tan \theta = -\frac{x_A + \delta x}{\hat y}.
    \label{eq:tantheta}
\end{equation}
The substituting of \eqref{eq:tantheta} into \eqref{eq:formula1} at order 1 in $\Delta n$, yields that
\begin{equation}
    \delta x=\left(y_A - \hat y \right)\frac{\Delta n x_A}{n_1\hat y}.
    \label{eq:formula2}
\end{equation}
The line (AP) thus satisfies that
\begin{align}\nonumber
    y - y_A &= \frac{\hat y - y_A}{\delta x} (x - x_A) \\ 
    &= -\frac{n_1 \hat y}{\Delta n x_A} (x - x_A).
\end{align}
The intersection for which $x=0$ corresponds to
\begin{equation}
    y_0 = y_A + \frac{n_1 \sqrt{R^2-x_A^2}}{\Delta n}.
\end{equation}
For $x_A \ll R$ (i.e., for points close to the optical axis, which corresponds to the paraxial approximation), we have 
\begin{equation}
    y_0 = y_A + \frac{n_1 R}{\Delta n}.
\end{equation}
The first term $y_A$ is negligible compared to the second term as $\Delta n \ll 1$ and $y_A\leqslant R$. 
This implies that the distance of the cone origin $O'$ to the origin of the coordinate system $O$ is much larger than the sample diameter.
We thus obtain that
\begin{equation}\label{eq:conecenter}
    y_0 = \frac{n_1 R}{\Delta n}.
\end{equation}

The coordinates of this intersection point $O'$ do not depend on the position of our point $A$ in the paraxial approximation. We thus obtain a cone-beam geometry where all the rays inside the cylinder intersect at $O'$. 
When $n_1=n_2$, the origin of the cone beam becomes $+\infty$; hence, the cone reduces to the standard parallel-beam model, as expected.

\section{Cone-Beam Shift of Origin Due to Off-Center Rotation}
\def\yy{y'}
\def\yc{y_0}

While we have shown that the index mismatch transforms the parallel problem into a conic one, the center that yields the optimal FDK reconstruction is not the $\yc$ predicted by \eqref{eq:conecenter}, but rather an $y_r$ that is significantly closer to the object, with $\yc>y_r$. 
Following recent evidence regarding the effects of recalibration on seagull artifacts \cite{Schmidt:21}, we have established experimentally that a mis\-centering of the cylinder with respect to the axis of the turning platform is at the origin of this offset. 

To reconcile the conic geometry motivated in \eqref{eq:conecenter} with this newly characterized experimental error, we proceed to show that the offset of the center of rotation accumulates the original projections in a way that generates an alternative cone with its center displaced towards the origin.
Indeed, even when misaligned, the platform still rotates the sample fully, and hence, generates complete tomographic data. 
The center of a cone beam corresponds to the position through which all light rays pass; we thus look for the point where the flux of intensity is maximal. 

\begin{figure}[htb]
    \centering
    \includegraphics[width=\linewidth]{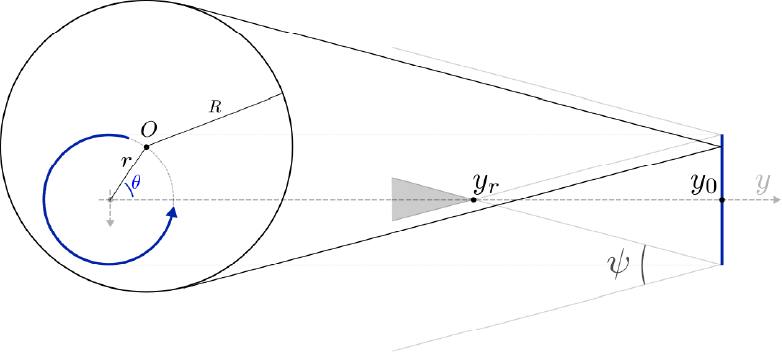}
    \caption{Effect of the off-centered rotation of the sample on the center of the cone beam. The bigger black disk corresponds to the cylinder in Fig. \ref{fig:geometry}, whereas the greyish inner circle represents its orbit around the true center (size exaggerated). The rotation trajectory (in blue) is translated into an oscillation (also in blue) of the cone's center that generates an alternative cone closer to the sample.}
\label{fig:offcenter}
\end{figure} 

The effect of the rotation on the $y$ direction being negligible because $y_0 \gg R$, we focus instead on the movement induced in the $x$ direction. 
This displacement can be expressed as $r\cos \theta$, where $\theta$ is the angle of rotation and $r$ the distance between the center of the sample and the true axis. 
At each angle of rotation, the rays originate a light cone that is viewed from $\yc$ with an angle $\psi$, which in turn depends on $R$ (Fig. \ref{fig:offcenter}). 
These cones accumulate on the line segment $[-r,r] \times \{\yc\}$ (blue in Fig. \ref{fig:offcenter}) according to the amplitude of rotation. 

Let us establish the coordinate system $(x,\yy)$ centered on $(0,\yc)$ and with the axis of symmetry flipped, so that $y=(\yc-\yy)$. 
In the $(x, y')$ system of coordinates, the finite domain limits the spatial influence of the rays, making them all coincide at $x=0$, $\yy=\yy^\star = r \cot \left( \psi /2 \right)$, and thus creating an equivalent cone center under fairly general conditions. 
To show this, we use $f(x,\yy)$ to model the brightness within the span of each cone and $w(x)$ to weight their contributions in a way that reflects how the projections concentrate at the poles. 
Together, they multiply to yield the weighted intensity flux
\begin{equation}
g (x, \yy) = f \left( \sqrt{x^2+\yy^2} \right) w(x).
\end{equation}
All contributions can then be combined by integrating over the $x$ movement as
\begin{equation}
I(\yy)/2 = \begin{dcases} 
      \int_0^r g\left( x, \yy \right) \,\text{d}x, & \yy \geq \yy^\star \\
      \int_0^{\yy\tan(\psi/2)} g\left( x, \yy \right)\,\text{d}x, & \yy<\yy^\star,
   \end{dcases}
\end{equation}
which is split into two components precisely because the offset rotation has finite amplitude.
Note that the weighted intensity energy $I$ is continuous as $\yy=\yy^\star$ is the solution to $r=\yy\tan(\psi/2)$, and that the factor of $2$ appears due to the symmetry of $g$ with respect to the integration domain.

The maximum of $I$ is attained at $\yy^\star$ when $g$ is constant, thus originating a (uniform) cone beam closer ($\yy^\star>0$) to the object.
We now show that such a maximum still exists under quite general physical conditions. 
On one hand, brightness is to be conserved, one expects the divergence of $f$ to vanish in the radial direction $\rho$ of the corresponding polar reference system, yielding that $\nabla \cdot f = \rho^{-1} \partial \left( \rho f \right) / \partial \rho=0$ and, thus, that $f \propto \rho^{-1} = 1/\sqrt{x^2+\yy^2}$. 
On the other hand, the weighting function can be thought to be proportional to the derivative of the geometric projection of the rotation, like $w\propto \left|\sin (\pi x/2r) \right|$. 
Together, $f$ and $w$ combine into a $g \propto \sin (\pi x/2r) / \sqrt{x^2+\yy^2}$ that makes $I$ monotonically increasing for $\yy<\yy^\star$ and decreasing for $\yy>\yy^\star$ because the integration domain shrinks progressively.
Therefore, with the maximum $\yy^\star=\text{argmax}_{\yy \in [0, \infty)} I$, this showcases the formation of an equivalent cone beam that is closer to the object. With respect to the original system of coordinates, the shifted cone-beam center is at $y_r=(\yc-\yy^\star)$. 

To find the center $y_r$ of the cone beam in practice, we resort to the automatic calibration scheme presented in the main text because it is more robust against seagull artifacts: it finds an origin that is consistent across several beads without the need to estimate additional experimental parameters such as $r$, $y_0$, or $\psi$. 

\end{document}